\documentclass[%
reprint,
superscriptaddress,
 amsmath,amssymb,
 aip,
]{revtex4-2}

\usepackage{graphicx}
\usepackage{dcolumn}
\usepackage{bm}
\usepackage{threeparttable}
\usepackage{mathrsfs}
\usepackage{tikz}

\begin{document}


\title{Femtosecond laser-induced sub-wavelength plasma inside dielectrics: III. Terahertz radiation emission}
\author{Kazem Ardaneh}
\email{kazem.arrdaneh@gmail.com}
\affiliation{FEMTO-ST Institute, Univ. Bourgogne Franche-Comt\'e, CNRS, 15B avenue des Montboucons, 25030, Besan\c{c}on Cedex, France}
\affiliation{Sorbonne University, Pierre and Marie Curie Campus, 4 place Jussieu, 75252, Paris Cedex 5, France}
\author{Ken-Ichi Nishikawa}
\affiliation{Department of Physics, Chemistry and Mathematics, V. Murry Chambers Bld., Alabama A\&M University, Huntsville, AL 35810, USA}
\author{Remo Giust}
\author{Benoit Morel}
\author{Pierre-Jean Charpin}
\affiliation{FEMTO-ST Institute, Univ. Bourgogne Franche-Comt\'e, CNRS, 15B avenue des Montboucons, 25030, Besan\c{c}on Cedex, France}
\author{Arnaud Couairon}
\affiliation{CPHT, CNRS, Ecole Polytechnique, Institut Polytechnique de Paris, Route de Saclay, F-91128 Palaiseau, France}
\author{Guy Bonnaud} 
\affiliation{CEA, Centre de Paris-Saclay, DRF, Univ.  Paris-Saclay, 91191 Gif-sur-Yvette, France}
\author{Francois Courvoisier}
\email{francois.courvoisier@femto-st.fr}
\affiliation{FEMTO-ST Institute, Univ. Bourgogne Franche-Comt\'e, CNRS, 15B avenue des Montboucons, 25030, Besan\c{c}on Cedex, France}

\date{\today}

\begin{abstract}
{
Electromagnetic radiation within the terahertz (THz) frequency range is of great interest for applications in remote sensing and time-domain spectroscopy.  The laser-induced plasmas are promising mediums for generating THz radiation. It has been recently reported that focusing femtosecond Bessel pulses inside dielectrics induces a high aspect ratio over-critical plasmas. Here we show that the intense resonantly driven electrostatic fields at the so-called critical surface lead to THz radiation emission.  Through three-dimensional particle-in-cell simulation and analytical derivation, we have investigated the emission of THz radiation. We show that the THz radiation is associated with a hot population of electrons trapped in ambipolar electric fields of the double layers.}
\end{abstract}

\maketitle

\section {Introduction}\label{Introduction}
Terahertz (THz) radiation, typically referred to as the frequency band 100 GHz $-$ 10 THz,  in the infrared and microwaves ranges,  has been attracting ongoing interest because of its broad applications ranging from biomedical imaging, security or packaged goods inspection,  to time-domain spectroscopy. \cite{Grischkowsky_90,amico_2008,Ferguson_2002, Masayoshi_2007}
Using femtosecond laser pulses, the techniques for generating THz radiation are generally classified as either optical rectification,\cite{Bass_1962,Xu_1992,Nahata_1996,Cook_2000,Kress_2004,Bartel_2005,Xie_2006} transient current sources,\cite{Auston_1998,Hamster_1993,Hamster_1994,Cheng_2001,Sprangle_2004,kim_2007,amico_2008,li_2012,li_2014,mitryukovskiy_2014,buccheri_2015,miao_2016,dechard_2020} or a combination of these two mechanisms. \cite{Andreeva_2016} 

Optical rectification can induce THz radiation in non-centrosymmetric crystals, e.g., ZnTe and GaAs, in which the fundamental frequency of an infrared femtosecond laser pulse is down-converted to the THz frequency via the second-order susceptibility where the polarization reads $\mathbf{P}\left(\omega_{\mathrm{THz}}\right)=\chi^{(2)} \mathbf{E}(\omega+\omega_{\mathrm{THz}}) \mathbf{E}^{*}(\omega)$. \cite{Bass_1962,Nahata_1996} The frequency of the rectified pulse envelope is in the range of 3-10 THz.\cite{Nahata_1996} In the centrosymmetric media, e.g., gases, two-color illumination can be used to mix the fundamental frequency with the second-harmonic in a four-wave mixing process as $\mathbf{P}\left(\omega_{\mathrm{THz}}\right)= \chi^{(3)} \mathbf{E}(2 \omega \left.-\omega_{\mathrm{THz}}\right) \mathbf{E}^{*}(\omega) \mathbf{E}^{*}(\omega)$. \cite{Cook_2000,Xie_2006} Correspondingly, the THz component might match with the fundamental one in an inverse process and leads to second-harmonic generation, which is a method for THz detection. A THz emission with an estimated field strength $\sim$ 400~kV/cm has been reported in which the presence of plasma was essential for the high-efficiency process. \cite{Bartel_2005}

Laser-induced plasmas are attractive for THz radiation generation because of  their ability to sustain extremely intense electromagnetic fields.\cite{Hamster_1993,Hamster_1994,Yoshii_1997,Leemans_2003,Sheng_2005} In this context, femtosecond laser-induced breakdown of gases is investigated widely,\cite{tzortzakis_2002,loffler_2005} mostly using the two-color approach.\cite{Cook_2000} The peak of the THz field however saturates for laser intensities higher than $10^{15}$~W/cm$^2$ because of the strong THz absorption in the long ($\sim$7~mm) air plasma.\cite{kim_2007} Plasma generation in laser-solid interactions offer an alternative: experiments of ultrashort laser pulses-solid interaction have shown a monotonous increase in THz radiation with the incident laser intensity up to $10^{19}$~W/cm$^2$.\cite{li_2012,li_2014,buccheri_2015}

Illumination of solid targets by intense ultrashort laser beams results in the generation of hot electron currents that are the source of the secondary electromagnetic radiation ranging from x-rays \cite{Teubner_1993,Sauerbrey_1994,Murnane_1989,Brambrink_2009} to THz radiation.\cite{li_2014} Under $p-$polarized laser illumination of a short-scale inhomogeneous plasma, for moderate laser intensities (bellow $10^{14}\, {\rm W/cm^2}$), resonance absorption is the main mechanism for hot electron generation.\cite{kruer_1988,eliezer_2002,gibbon_2005}  

\begin{figure*}[!htp]
\begin{center}
\includegraphics[width=\textwidth]{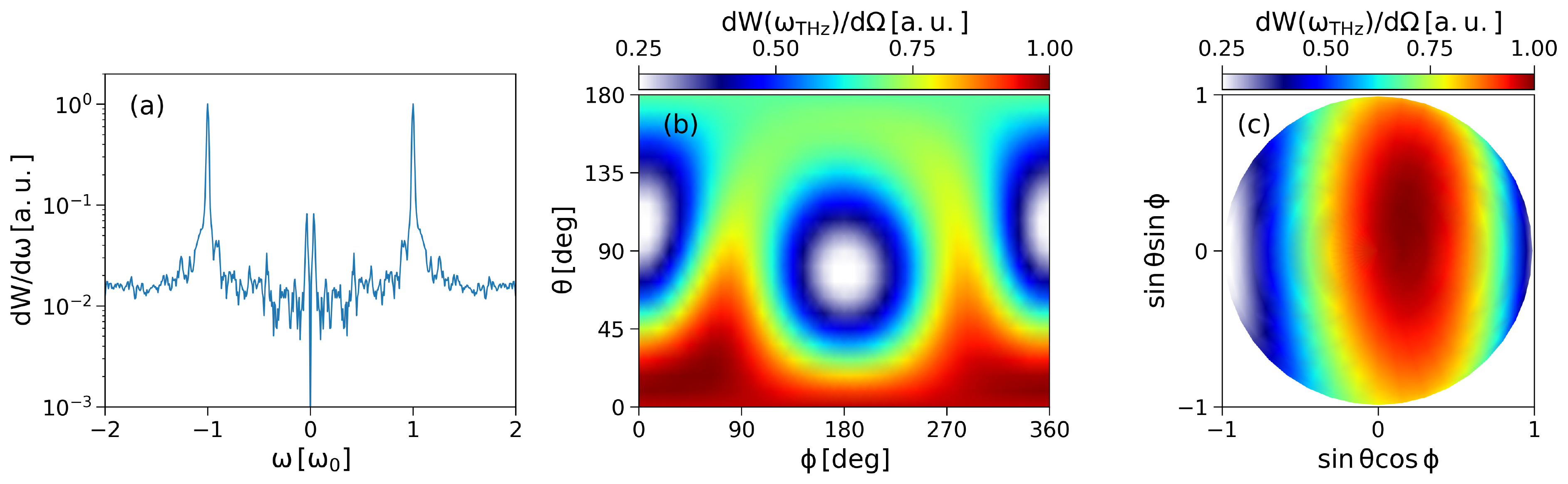}
\caption{The energy spectrum of the radiation emitted by a population of hot electrons: (a) the spatially averaged, (b) angular distribution, and (c) spatial distribution averaged in the frequency range of 1-30 rad/ps. The hot electrons are randomly selected.}
\label{energy_spe_fig}
\end{center}
\end{figure*}

{In previous papers, we have reported that over-critical plasmas, i.e. with density above the reflection density for the incident laser wavelength, are generated by focusing Bessel beams with moderate intensities on the order of $10^{14}\, {\rm W/cm^2}$ inside sapphire.}\cite{ardaneh_2021,Ardaneh_2022a,Ardaneh_2022b} A Bessel beam is a solution of the wave equation in which the wave amplitude is defined by the Bessel function of the first kind.\cite{Durnin87} Importantly, the axial intensity profile of the Bessel beam is propagation invariant. Therefore, all segments of the dielectric along the Bessel zone will receive simultaneously the same amount of energy which results in a high aspect ratio plasma rod. 

In the first article of this series (Ardaneh {\it et al.},\cite{Ardaneh_2022a} Paper I hereafter), we confirmed that the resonance of the plasma waves can explain the experimental diagnostics of total absorption, and far-field intensity pattern. We reported electron acceleration up to several keV while surfing the plasma waves. In the outward propagation of hot electrons, electrostatic ambipolar fields form at the plasma surface due to the different inertia of the electron and ion. Moreover, in the second article of this series (Ardaneh {\it et al.},\cite{Ardaneh_2022b} Paper II hereafter), we reported the second-harmonic generation by a second-order current of hot electrons near the critical surface. The electron currents form by the resonance absorption and radiation force of the incident laser wave. 

{In the current work as the third in this series, we establish a link between resonance absorption-driven currents and THz radiation. This is based on calculating the coherent radiation spectrum of the hot electrons for the performed Particle-In-Cell (PIC) simulation. The simulation consists of electron-ion plasma initially induced by multi-photon and collisional ionizations. The dipole moments are induced due to the radiation force of the resonance fields. For a laser field with frequency $\omega_0$, this force induces a second-harmonic component at $2\omega_0$ and a low-frequency component by separating the light electrons from the heavy ions.  We have developed an analytical model for THz generation in laser-plasma interactions to explain the underlying physics, in particular, how the dipole moment is created in the plasma and the characteristics of the radiation spectrum. }

We organized the paper as follows. In Sec. \ref{PIC Simulation}, we recall the setup of the PIC simulation as discussed in Paper I\cite{Ardaneh_2022a}, we detail the radiation diagnostic, and the results of the simulation. Then, in Sec. \ref{Electron THz emission in ambipolar electric fields}, {we derive an analytical solution for the current source of THz radiation,  and the radiated electromagnetic fields with their frequency spectrums. }

\begin{table}[!htb]
\centering
\begin{threeparttable}
\caption{ {Simulation setup}.\label{dsh}}
\begin{tabular}{ |p{4cm}||p{4cm}|  }
 \hline
Parameter & Value \\
 \hline
Simulation volume    & $15\times15\times 30$~$\rm \mu m^3$   \\
Grid resolution     &   $\Delta_{\rm x:y}k^{0}_{\rm r}=0.04$\tnote{a},  $\Delta_{\rm z}k^{0}_{\rm z}=0.1$\tnote{b} \\
FDTD order  & Second-order  \\
BC for fields\tnote{c} &  Perfectly matched layers\\
BC for particles &  Outflow\\
Pulse energy ($E_{\rm p}$)&1.2 $\rm \mu J$ \\
Pulse frequency ($\omega_0$)  & 2.35~rad/fs\\
Pulse cone angle ($\theta$)  & $25 ^{\circ}$\\
Pulse temporal profile & $\exp[-(t-t_{\rm c})^2/T^2]$\\
Central time ($t_{\rm c}$) & $130\,{\rm fs}$ \\
Pulse ${\rm FWHM}=\sqrt{2\ln2}T$    & $100\,{\rm fs}$\\
Pulse spatial profile &  $\exp(-r^2/w_0^2)$ \\
Pulse spatial waist ($w_0$) & 10~$\rm \mu m$ \\
Maximum density ($n_{{\max}}$) & $5\,n_{\rm c}$\\
Density profile  (axial) & $\tanh(z^{\rm {\mu m}})$\\
Mass ratio ($m_{\rm i}/m_{\rm e}$) & $102\times1836$\tnote{d}\\
Plasma distribution [$f(\mathbf{v}_{\rm e:i})$]  & Maxwellian\\
Plasma temperature ($T_{\rm e:i}$) & $1$~eV\\
Particles per cell per species &  32\\
Particle shape function&  triangle\\
Time step & $\Delta t \omega_0= 0.07$ \\
Simulation time & $320\,{\rm fs}$\\
 \hline
\end{tabular}
\begin{tablenotes}
\item [a] $k^{0}_{\rm r}=k_0\sin\theta$. \item [b] $k^{0}_{\rm z}=k_0\cos\theta$. \item [c] BC: Boundary condition. \item [d] 102 is the sapphire molar mass. 
\end{tablenotes}
\end{threeparttable}
\end{table}

\section {PIC Simulation} \label {PIC Simulation}
 {
We performed self-consistent PIC simulation using the three-dimensional massively parallel electromagnetic code EPOCH \cite{Arber_2015}. In our simulation, we used the plasma parameters that could reproduce our experimental measurements (far-field, near-field, absorption) as reported in Paper I,\cite{Ardaneh_2022a} and II.\cite{Ardaneh_2022b} 
The simulation setup is summarized in Table \ref{dsh}. The plasma is fully ionized and composed of electrons and ions with equal densities (to preserve electric neutrality) given by $n =n_{\max}\exp(-x^2/w_{\rm x}^2)\exp(-y^2/w_{\rm y}^2)\tanh(z^{\rm {\mu m}})$ with FWHM$_{\rm x}=\sqrt{2 \ln 2}w_{\rm x}=250$~nm, and FWHM$_{\rm y}=\sqrt{2 \ln 2}w_{\rm y}=600$~nm.  There are initially 32 particles per cell per species leading to the total number of particles in the simulation $\sim10^9$. The collisions are modeled through a binary model as presented in Refs.\cite{SENTOKU_2008,Arber_2015}. 

We injected from the $z_{\min}$ boundary a linearly $x-$polarized Gaussian pulse propagating along the positive $z-$direction.  We applied a phase to the Gaussian beam to create a Bessel-Gauss beam.\cite{Ardaneh_20} The peak intensity in the Bessel zone is $6\times 10^{14}\,{\rm W/cm^{2}}$ in absence of plasma. The time step is limited by the Courant condition.  The minimum frequency in the simulation is 1.5~rad/ps which is well below the peak frequency of the THz spectrum at 30~rad/ps. 
}

\begin{figure*}[!htp]
\begin{center}
\includegraphics[width=\textwidth]{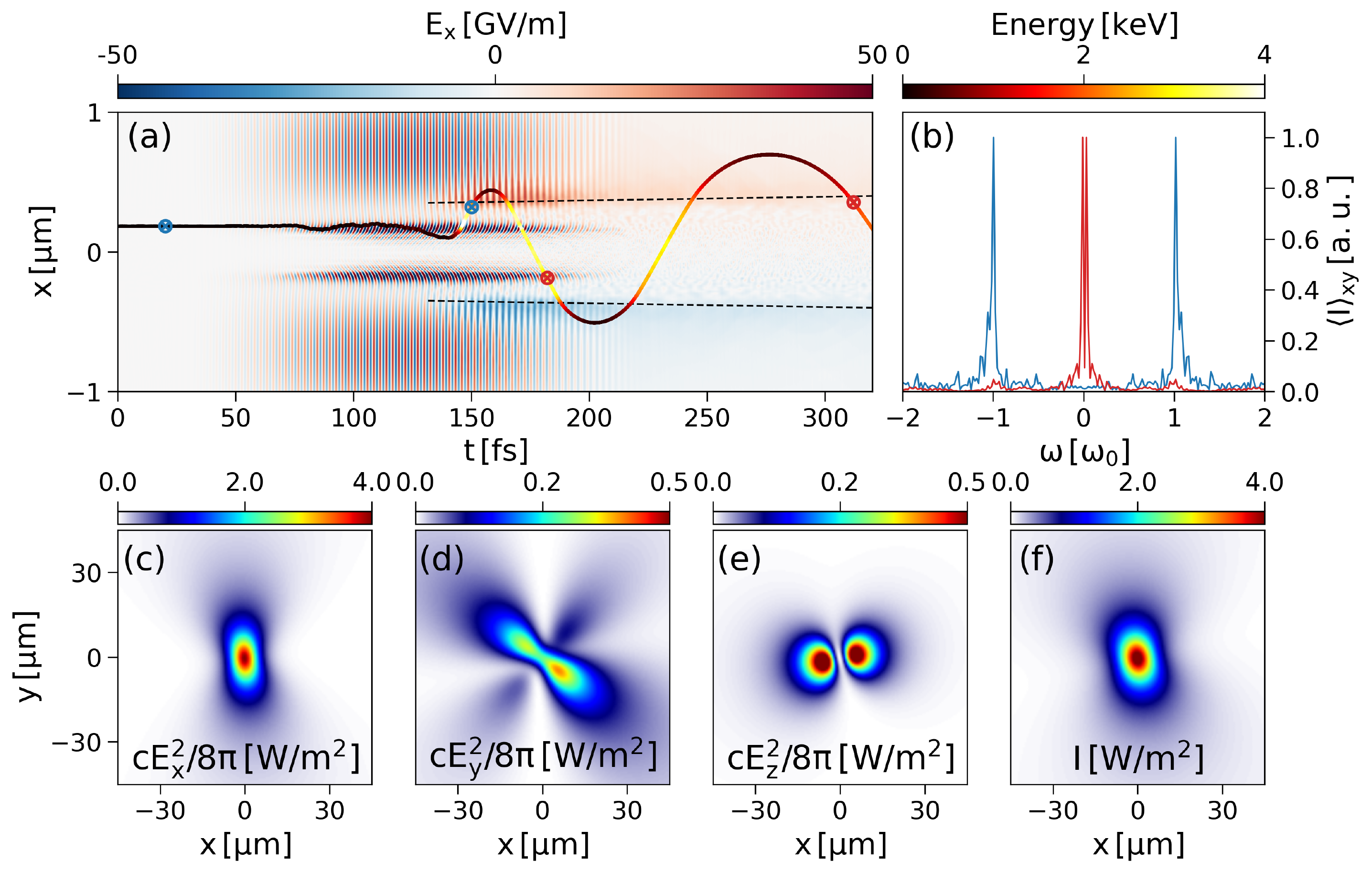}
\caption{THz radiation from an electron trapped in the ambipolar electric fields of double layers. Shown are: (a) $x-$component of the electric field over-plotted by the trajectory of a representative electron, electron emission for a time window between: (b) 20-150~fs [shown by blue $\bigotimes$ symbols in panel (a), and blue line in panel (b)], and 182-312~fs [shown by red $\bigotimes$ symbols in panel (a) and red line in panel (b)], (c-f) { $cE^2_{\rm x}/8\pi$, $cE^2_{\rm y}/8\pi$, $cE^2_{\rm z}/8\pi$ and the total intensity of the THz radiation emitted by the electron for the time window between 182-224~fs. The dashed lines in panel (a) show the expansion of the plasma at the sound velocity. The color in the electron trajectory reflects its energy based on the color bar of the panel (b).} }
\label{pow_spe_fig}
\end{center}
\end{figure*}

One of the primary advantages of PIC codes is the possibility to access full information about the particles. We have developed a radiation diagnostic that utilizes the position and momentum of particles over time and calculates the radiated fields and energy. For this purpose, let us consider a particle at position ${\bf r}\left(t\right)$ at time $t$. At the same time, we observe the radiated electromagnetic fields from the particle at position ${\bf x}$.  Due to the finite velocity of light, we observe the particle at an earlier position ${\bf r}\left(t^{\prime}\right)$ where it was at the retarded time $t^{\prime}=t-\mathbf{R}\left(t^{\prime}\right) / c$, where $\mathbf{R}\left(t^{\prime}\right)=\left|\mathbf{x}-\mathbf{r}\left(t^{\prime}\right)\right|$ is the distance from the charged particle (at the retarded time $t^{\prime}$) to the observer. The magnetic and electric fields produced from a moving point charge can be calculated directly from their scalar and vector potentials known as the Li\'{e}nard–Wiechert potentials. The electric field reads:\cite{jackson_1998}

\begin{equation}\label{radiation_EB}
\begin{split}
\mathbf{E}(\mathbf{x}, t)=&\overbrace{e\left[\frac{\mathbf{n}-\boldsymbol{\beta}}{\gamma^{2}(1-\boldsymbol{\beta} \cdot \mathbf{n})^{3} R^{2}}\right]_{\mathrm{ret}}}^{\text{Velocity field}}\\
+&\overbrace{\frac{e}{c}\left[\frac{\mathbf{n} \times\{(\mathbf{n}-\boldsymbol{\beta}) \times \dot{\boldsymbol{\beta}}\}}{(1-\boldsymbol{\beta} \cdot \mathbf{n})^{3} R}\right]_{\mathrm{ret}}}^{\text{Acceleration field}}
\end{split}
\end{equation}
where $\mathbf{n}=\mathbf{R}\left(t^{\prime}\right) /\left|\mathbf{R}\left(t^{\prime}\right)\right|$ is a unit vector pointing from the particle retarded position to the observer, $\boldsymbol{\beta}=\mathbf{v}/c$ the particle instantaneous velocity, $\dot{\boldsymbol{\beta}}={\rm d} \boldsymbol{\beta} / {\rm d} t$ is the acceleration divided by $c$, $\gamma$ is the Lorentz factor. The spatial spectra are obtained by the choice of $\mathbf{n}\left(n_{\rm x}^{2}+n_{\rm y}^{2}+n_{\rm z}^{2}=1\right)$. The field in Eq. (\ref{radiation_EB}) divides itself into "velocity fields," which are independent of acceleration, and "acceleration fields," which depend linearly on $\dot{\boldsymbol{\beta}}$. The velocity field is a static field decreasing as $R^{-2}$ while the acceleration field is a radiation field, being transverse to the radius vector and falling off as $R^{-1}$. The total energy $W$ radiated per unit solid angle ${\rm d} \Omega$ per unit frequency ${\rm d} \omega$ from the accelerated charged particle reads:\cite{jackson_1998}

\begin{equation}\label{energy_spe}
\frac{{\rm d}^{2} W}{{\rm d} \omega {\rm d} \Omega}=\frac{e^{2}\omega^2}{4 \pi^{2} c}\left\vert\int_{-\infty}^{\infty}  {\rm d}t^{\prime} {\hat{\mathbf{n}}\times(\hat{\mathbf{n}}\times\boldsymbol{\beta})} e^{j \omega(t^{\prime}+R(t^{\prime}) / c)} \right\vert^{2}
\end{equation}

In our simulations, we collected the THz radiation emissions from the hot electrons as follows. We have tracked $10^5$ electrons in the simulations and recorded the information of these electrons. We calculated the energy spectrum of the radiation emitted by 100 randomly selected electrons according to Eq. (\ref{energy_spe}). The result is shown in Fig. \ref{energy_spe_fig}(a). We see two peaks around $\omega=0$, with a width of typically 50 rad/ps. We note that, because of computing memory limitations, the time resolution of the particle positions is insufficient to capture the second harmonic emission. The angular and spatial distributions of the THz emission are obtained by averaging the energy spectrum in the frequency range of 1-30 rad/ps [Figs. \ref{energy_spe_fig}(b) and \ref{energy_spe_fig}(c)]. 

As one expects, the energy spectrum has sharp maxima at the laser frequency $\omega_0$ due to strong electron acceleration in resonantly driven plasma waves at the critical surfaces. The spectrum also has maxima at $\omega\approx30$~rad/ps. The angular distribution of this THz radiation in Fig. \ref{energy_spe_fig}(b) shows maxima around $(\theta, \phi)\approx(0, \pi/2)$ and $(0, 3\pi/2)$, perpendicular to the electron acceleration which is mainly in the $x-$direction. The small tilt in Fig. \ref{energy_spe_fig}(c) is due to the asymmetric distribution of the randomly selected electrons in $xy-$space over the integrated time (a similar deviation occurred for another set of 100 electrons). 

We select some representative electrons to calculate the radiated fields in a spatial window $\vert x \vert \leqslant 45$~$\rm{\mu m}$  and $\vert y \vert \leqslant 45$~$\rm{\mu m}$  at $z=0$. Using a time window, we also examined in which part of the trajectory, the electron emits electromagnetic radiation in the THz frequency range [Figs. \ref{pow_spe_fig}(a)]. For each time window, we calculated the intensity distribution $I(\omega_{\rm THz}, x, y)$ by performing a discrete Fourier transform on each component of the electric field, $E_{\rm x:y:z}(t,x,y)$ and averaging in the frequency range of 1-30 rad/ps [Figs. \ref{pow_spe_fig}(b)-\ref{pow_spe_fig}(f)]. 

Figure \ref{pow_spe_fig}(a) shows the time evolution of the $E_{\rm x}$ component, parallel to the incident laser polarization over-plotted with the trajectory of a representative electron. {One can see the resonance plasma waves induced at the critical surfaces ($x=\pm0.2\, {\rm \mu m}$), in the time between 70-200~fs (See Fig. 4 in Paper I\cite{Ardaneh_2022a} for more details). Near the peak of the laser field, the intense ambipolar fields propagating with the sound speed are generated at the surface of the plasma (dashed lines).  The ambipolar field sign is positive for $x > 0$ and negative for $x < 0$. The radiation force (See Appendix \ref{Radiation force density}) due to the intense, localized resonance field ejects electrons from the resonance region. The electrons are ejected from the critical surfaces in the positive $x-$direction where $x>0$ and negative $x-$direction where $x<0$ as shown in Fig. 6 Paper I. Therefore, the electrons ejected with energy less than the potential barrier of the ambipolar field will be reflected by $-e\bf{E}$ force.  Consequently,  these electrons will be trapped between the ambipolar electric fields on either side of the plasma. An ejected electron oscillates between the ambipolar fields with a period that increases with time due to the energy exchange between the electrons and ions. } 

We monitored the electron radiation using a time window of 130~fs. The electron emission between 20-150~fs [shown by blue $\bigotimes$ symbols in panel (a)], is sharply peaked at the laser frequency $\omega_0$ [blue line in panel (b)]. This emission is due to the electron acceleration while surfing the resonantly driven plasma waves (See Paper I\cite{Ardaneh_2022a} for more details). {During the time interval 182-312~fs [shown by red $\bigotimes$ symbols in panel (a)], the electron is trapped between two ambipolar fields and emits the THz radiation with a peak frequency at $\omega=24$~rad/ps [red line in panel (b)]. Figures \ref{pow_spe_fig}(c-f) show respectively the intensity of the electric field components computed using Eq. (\ref{radiation_EB}), $c(E^2_{\rm x}, E^2_{\rm y}, E^2_{\rm z})/8\pi$, and the total intensity radiated by the electron during the time between 182-312~fs. The THz radiation is mainly polarized in the $x-$direction because $E_{\rm x}$ component is dominant in the radiated field. This polarization is the same as the incident pulse and second-harmonic detailed in Paper II\cite{Ardaneh_2022b}.}
 In Sec. \ref{Electron THz emission in ambipolar electric fields}, we will show that the emission pattern corresponds with an electron current in the $x-$direction. 

\section {Electron THz emission in ambipolar electric fields} \label{Electron THz emission in ambipolar electric fields}
The starting point in understanding the mechanism responsible for THz radiation is the identification of its current sources. We have seen earlier, in Figs. \ref{pow_spe_fig}, that the electrons emit THz radiation while they are trapped in the ambipolar electric fields of  { plasma double layers}. Taking the strength of the resonance electric field of about 50~GV/m integrated over its width of 70 nm (See Paper I\cite{Ardaneh_2022a}), one arrives at a potential of about a few keV which corresponds to the temperature of the hottest electrons in the simulation. The hot electrons propagate outside of the plasma and consequently, the separation of charges forms an electric double layer where an ambipolar electric field is present. An analytical solution for this field is possible by using the two-fluid plasma equations for continuity and momentum (See Appendix \ref{Ambipolar electric field of double layer}).

Here for simplicity, we considered a $s-$polarized monochromatic laser wave as $\mathbf{E}=\mathbf{E}_{\rm s}(\mathbf{r}) \cos (\omega_0t)$ where $\mathbf{E}_{\rm s}(\mathbf{r})$ includes the spatial dependence. The ambipolar electric field $E_{\rm a}$ is described by an inhomogeneous second-order differential equation for a classical, damped, driven harmonic oscillator given by [See Eq. (\ref{2nd-inhomo}) in Appendix \ref{Ambipolar electric field of double layer}]: 

\begin{equation}\label{2nd-inhomo_main_tex}
    {\partial_{\rm t}^{2} E_{\rm a}}+2\Gamma {\partial_{\rm t} E_{\rm a}}+\Omega_{\rm p}^2E_{\rm a}=\Omega_{\rm p}^2\left[\mathscr{E}_{0}+\mathscr{E}_{2}\cos(2\omega_0t)\right]
\end{equation}
where $\Gamma={\nu_{\rm ei}}/{2}\left(1+{Z m_{\rm e}}/{m_{\rm i}}\right)$, $\nu_{\rm ei}$ is the electron-ion collision frequency, $\Omega_{\rm p}^2=\omega_{\rm{pe}}^2\left(1+{Z m_{\rm e}}/{m_{\rm
i}}\right)$, $\omega_{\rm pe}=(4\pi n_{\rm e} e^2/m_{\rm e})^{1/2}$ is the electron plasma frequency (in cgs units), and

\begin{subequations}\label{cons_main_tex}
\begin{align}
\begin{split}
\mathscr{E}_{0}=&\frac{4 \pi e}{\Omega_{\rm p}^2}\left[{\partial_{\rm x}}\left(Z\frac{P_{\rm i}}{m_{\rm i}}-\frac{P_{\rm e}}{m_{\rm e}} + Zn_{\rm i} v_{\rm i}^{2}-n_{\rm e} v_{\rm e}^{2}\right)\right]\\
&-\frac{4 \pi e}{m_{\rm e}{\Omega_{\rm p}^2}}  \frac{\omega_{\rm{pe}}^{2}}{\omega_0^{2}}\partial_{\rm x}\left\langle\frac{{E}^2}{8\pi}\right\rangle
\end{split}\\
\mathscr{E}_{2} = & -\frac{4 \pi e}{m_{\rm e}{\Omega_{\rm p}^2}}  \frac{\omega_{\rm{pe}}^{2}}{\omega_0^{2}}\partial_{\rm x}\left\langle\frac{{E}^2}{8\pi}\right\rangle
\end{align}
\end{subequations}
with the standard notation $(t, x, v, P, m_{\rm s}, n_{\rm s}, Z)$ for the time, space, velocity, pressure, mass and density of a particle of species $s$, and ion charge respectively. The $\langle\rangle$ denotes an average over a laser cycle. The coupling to the laser was included in the momentum equation via the radiation force density ${\bf f}_{\rm RF}=(\epsilon-1)/{8 \pi} \boldsymbol{\nabla} E^{2}$ where $\epsilon$ is the plasma permittivity (See Appendix \ref{Radiation force density}). One can find an equation similar to Eq. (\ref{2nd-inhomo_main_tex})  in Refs.\cite{Sprangle_2004,amico_2008,mitryukovskiy_2014,Andreeva_2016} but with a different right-hand side (different current sources of the THz radiation). 

The solution of Eq. (\ref{2nd-inhomo_main_tex}) under the initial conditions of $\left(E_{\rm a},\partial_{\rm t}E_{\rm a}\right)=\left(0,0\right)$ reads (See for example Ref.\cite{Mathematica_13.1}):

\begin{figure*}[!htp]
\begin{center}
\includegraphics[width=\textwidth]{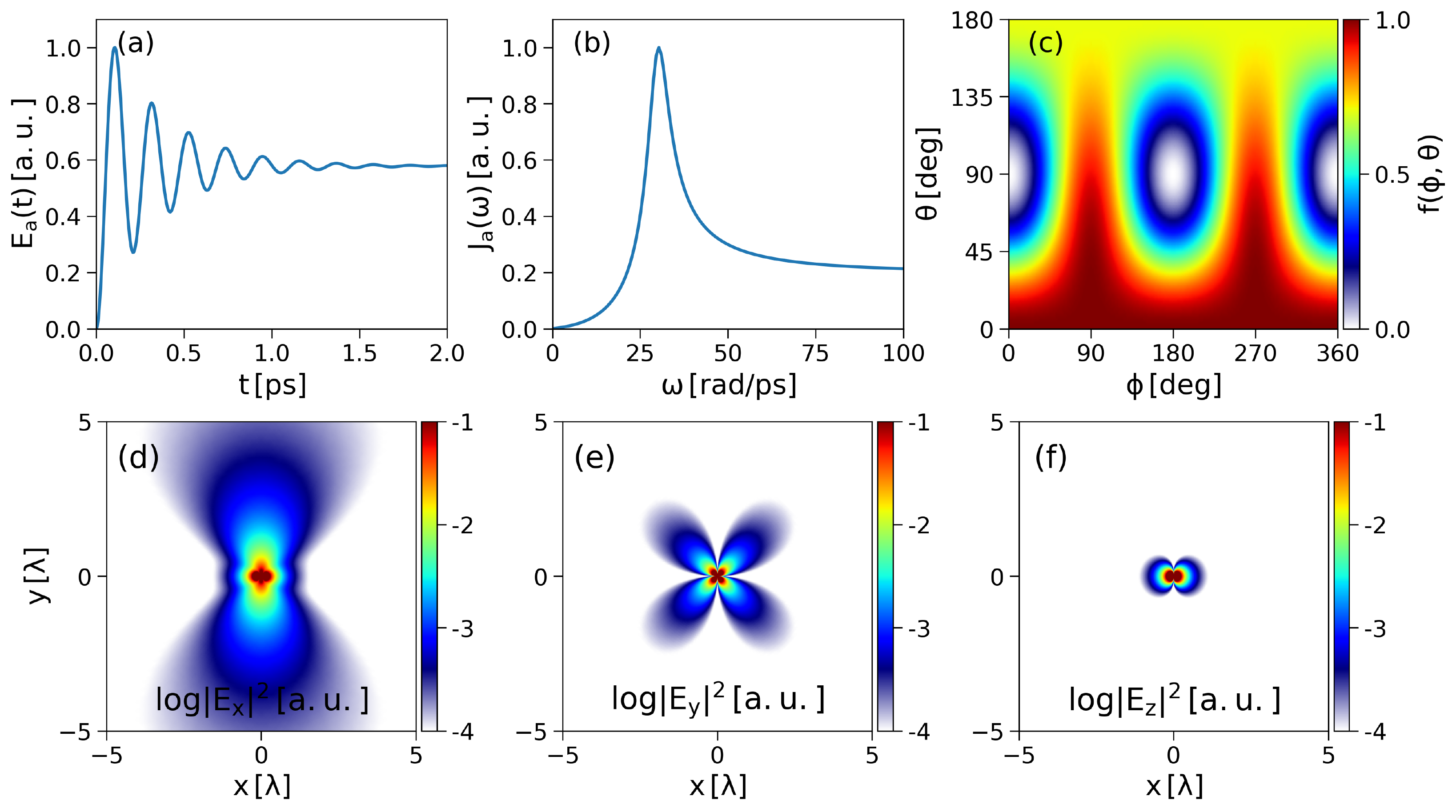}
\caption{Temporal profile of the THz component of the ambipolar electric field from Eq. \ref{solution}, panel (a), the frequency spectrum of the electromagnetic radiation, panel (b),  angular distributions of the radiated energy from Eq. \ref{ana_spe_ana}(b), panel (c), and electric field components, panels (d)-(f).}
\label{dl_fields}
\end{center}
\end{figure*}

\begin{equation}\label{solution}
    \begin{split}
    E_{\rm a}(t)=&\overbrace{\mathscr{E}_{0}\left\{1-\exp{\left(-\Gamma t\right)}\left[\cos\left(\varpi t\right) + \frac{\Gamma}{\varpi}\sin\left(\varpi t\right) \right]\right\}}^{\text{Terahertz oscillation}}\\
    +&\overbrace{\Omega_{\rm p}^2\mathscr{E}_{2}\frac{\left(\Omega_{\rm p}^2-4\omega_0^{2}\right)\cos(2\omega_0t) + 4\omega_0\Gamma\sin(2\omega_0t)}{\left(\Omega_{\rm p}^2-4\omega_0^2\right)^2+16\Gamma^2\omega_0^{2}}}^{\text{Second-harmonic oscillation}}
\end{split}
\end{equation}

where $\varpi^{2}=\Omega_{\rm p}^2-\Gamma^{2}$. This solution includes two components. The first component oscillates with a frequency close to the plasma frequency $\omega_{\rm{pe}}$ when $\omega_{\rm{pe}}\gg\nu_{\rm ei}$. This oscillation, however, decays exponentially at a rate close to the collision frequency. {This component is established by the spatial gradients of the pressure difference between the light electrons and the heavy ions as represented in Eq. (\ref{cons_main_tex}a).} This part induces the dipole moment in the plasma by separating the electrons from the ions. After a time $t\gg1/\nu_{\rm ei}$, neglecting the electron and ion velocities and assuming $T_{\rm e}\gg T_{\rm i}$ (See Paper I\cite{Ardaneh_2022a}), a nearly constant electric field remains $eE_{\rm a}\approx -{1}/{n_{\rm e}}{{\rm d}P_{\rm e}}/{{\rm d} x}=-\gamma T_{\rm e}{{\rm d} \ln n_{\rm e}}/{{\rm d} x}$, considering an adiabatic equation of state with the adiabatic index $\gamma$. Therefore, the ambipolar field oscillations are driven by the electron density gradient. The work function of the electrons that moved from the plasma interior (density $n_1$) to the exterior (density $n_2$) then reads $-e\Delta\phi=\gamma T_{\rm e}\ln(n_1/n_2)\approx4\,{\rm keV}$. 

The second part in Eq. \ref{solution} arises where gradients of the laser intensity induce a second harmonic longitudinal field oscillation. This term has a resonance at $2\omega=\Omega_{\rm p}\approx \omega_{\rm pe}$ (four times the critical density) for the evanescent part of the wave  causing a very steep increase of the oscillation amplitude. This resonance for $s-$polarized lasers is different from the Denisov resonance absorption occurring under oblique incidence of $p-$polarized lasers. \cite{goldsworthy_1986,hora_2008}

 { An example of the electric field given by Eq. (\ref{solution}) is shown in Fig. \ref{dl_fields}(a) for $\varpi=30$~rad/ps (corresponds to an edge plasma density of $n_2/n_{\rm c}=10^{-4}$), and $\Gamma=3$~rad/ps where we have supposed that the electron collision frequency is small compared to the plasma frequency. The  wave shows 2-3 oscillations and damps out within a time scale of $\sim$ 2~ps. The Fourier spectrum of the electric current associated with the quasi-static electric field, $4\pi J_{\rm a}(\omega)= j\omega E_{\rm a}(\omega)/(1+j\nu_{\rm ei}/\omega)$, is shown in Fig. \ref{dl_fields}(b). As one can see, it has a maximum at $\omega\approx\varpi=30$~rad/ps. 

To compare with the numerical results of the radiated emission in Fig. \ref{energy_spe_fig}(b), we derive the angular distribution of the radiated energy from a current source. For a number of accelerated charges, the integrand in Eq. (\ref{energy_spe}) involves the replacement $e \boldsymbol{\beta} e^{j\omega R(t^{\prime})/ c} \rightarrow \sum_{\rm m=1}^N e_{\rm m} \boldsymbol{\beta}_{\rm m} e^{j\omega R_{\rm m}(t^{\prime})/ c}$. In the limit of a continuous distribution of charge, the summation becomes an integral over the current density as $e\boldsymbol{\beta} e^{j\omega R(t^{\prime})/ c} \rightarrow 1/c\int {\rm d}^{3} {\bf r}^{\prime}\  \mathbf{J}({\bf r}^{\prime}, t^{\prime}) e^{j \omega  R(t^{\prime})/c}$.
Hence, the radiation energy per solid angle per frequency of the current source reads:\cite{jackson_1998}}

\begin{equation}\label{j_x_int}
\frac{{\rm d}^{2} W}{{\rm d} \omega {\rm d} \Omega}=\frac{\omega^{2}}{4 \pi^{2} c^{3}}\left|\int {\rm d} t^{\prime} \int {\rm d}^{3} {\bf r}^{\prime}\ {\hat{\mathbf{n}}} \times\left[{\hat{\mathbf{n}}} \times \mathbf{J}({\bf r}^{\prime}, t^{\prime})\right] e^{j \omega[t^{\prime}+R(t^{\prime})/c]}\right|^{2}
\end{equation}

We consider an emission length of $L$ for the plasma rod oriented parallel to the $z-$direction, and a current of electron in the $x-$direction as ${\bf J}({\bf r}^{\prime}, t^{\prime})={\hat{\mathbf{x}}}J_{\rm a}(t^{\prime})\delta(x^{\prime}) \delta(y^{\prime})\exp{\left(jkz^{\prime}\right)}$. For a coordinate system with the spherical angle $\theta=\cos ^{-1}\left({z}/{r}\right)$ and the azimuth angle $\phi=\tan ^{-1}\left({y}/{x}\right)$ defining the direction of observation $\mathbf{n}$, Eq. (\ref{j_x_int}) reduces to: 

{
\begin{subequations}\label{ana_spe_ana}
\begin{align}
    \frac{\mathrm{d}^{2} W}{\mathrm{d} \omega \mathrm{d} \Omega}=&\frac{\left|J_{\rm a}(\omega)\right|^{2}}{\pi^2 c}  f( \phi,\theta )\\
    f( \phi,\theta )=&\frac{\sin ^{2}\theta \sin ^{2}\phi + \cos ^{2}\theta}{\left(1-\cos\theta\right)^2} \sin ^{2}\left(Lk\sin^{2}\frac{\theta}{2}\right)
    \end{align}
\end{subequations}
}
where $k$ is the emission wave-vector. 

The angular distribution given in Eq. (\ref{ana_spe_ana}b) is shown in Fig. \ref{dl_fields}(c) for an emission length of $L=10\,{\rm \mu m}$. It fits well with the distribution obtained from the PIC simulation in Fig. \ref{energy_spe_fig}(b) where the emission is beamed in the positive $z-$direction and has two maxima in the $y-$direction, perpendicular to the current source. The vector potential for this current source is in the $x-$direction and at far-field, it reads:\cite{jackson_1998}

\begin{equation}
\begin{split}
 \mathbf{A}(\mathbf{x},\omega)=&\frac{1}{c} \frac{e^{j k R}}{R} \int {\rm d}^{3} \mathbf{r}^{\prime} \mathbf{J}\left(\mathbf{r}^{\prime},\omega\right) e^{-j k \mathbf{n} \cdot \mathbf{r}^{\prime}} \\
    =&{\hat{\mathbf{x}}}\frac{2J_{\rm a}(\omega)}{c} \frac{e^{j k R}}{kR}\frac{\sin\left(Lk\sin^{2}\frac{\theta}{2}\right)}{1-\cos\theta}
\end{split}
\end{equation}
    
One can derive the scalar potential $\Phi$ using the Lorenz gauge and then the components of the electric field as follows:

\begin{subequations}\label{ana_em_fileds}
\begin{align}
    \boldsymbol{\nabla} \cdot \mathbf{A}=&-\frac{1}{c} \frac{\partial \Phi}{\partial t}\\
\mathbf{E}=&-\frac{1}{c}\frac{\partial \mathbf{A}}{\partial t}-\boldsymbol{\nabla} \Phi
\end{align}
\end{subequations}

The components of the radiated electric field calculated using Eq. (\ref{ana_em_fileds}b) are shown in Fig. \ref{dl_fields}(d)-(f). In agreement with the results of the PIC simulation [Figs. \ref{pow_spe_fig}(c-d)], the radiated emission is polarized in the $x-$direction. Moreover, the angular distributions of the electric field components agree with the results of the PIC simulation [Figs. \ref{pow_spe_fig}(c)-(e)]. {The quadrupole pattern in Fig. \ref{pow_spe_fig}(d) is not symmetric like Fig. \ref{dl_fields}(e). The asymmetry is because the trajectory of the electron is not symmetric as the electron spends more time in the $x>0$ region relative to $x<0$ [Fig. \ref{pow_spe_fig}(a)].}

\begin{figure}[!htp]
\begin{center}
\includegraphics[width=\columnwidth]{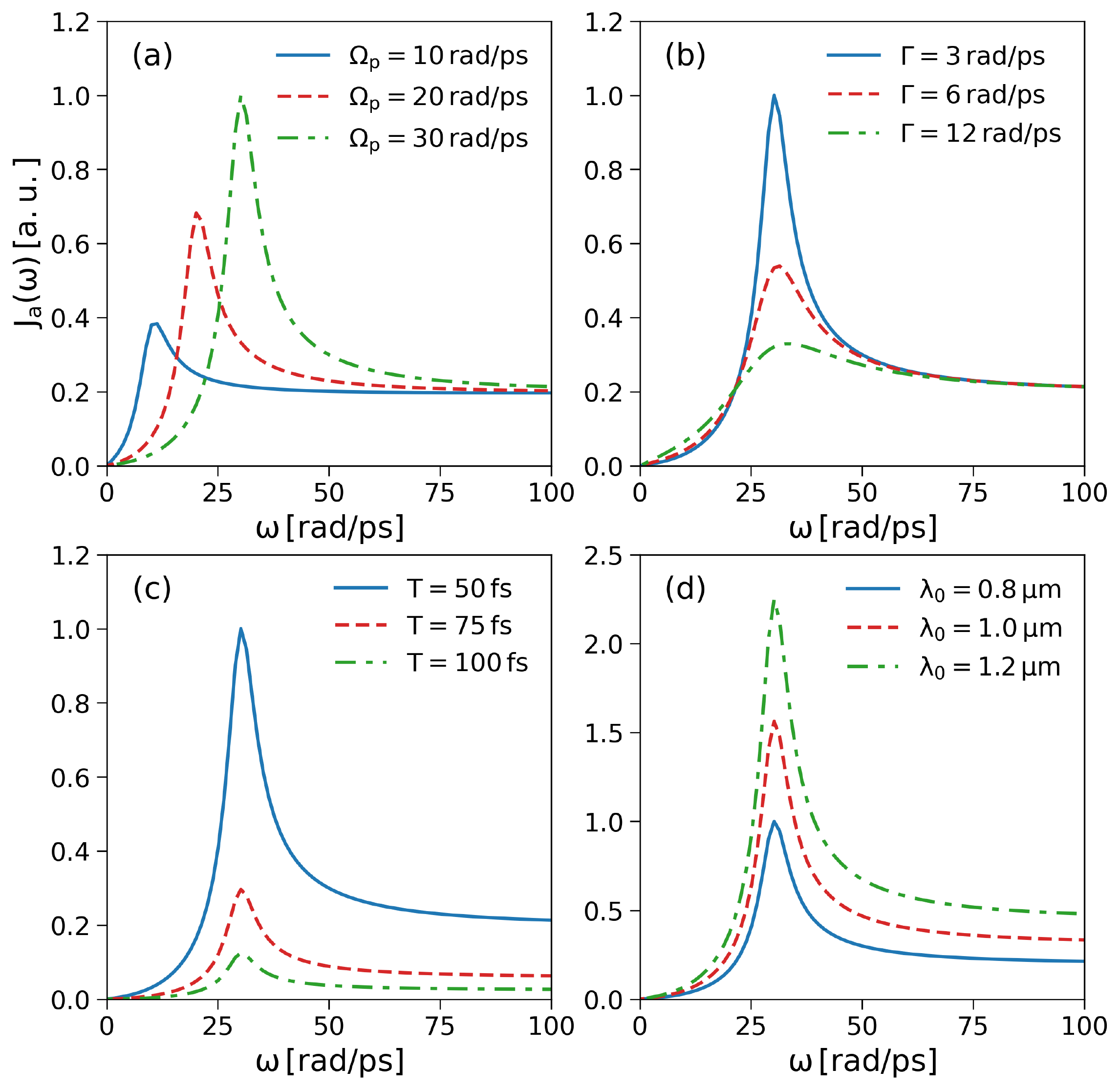}
\caption{ {  The frequency spectrum of the THz wave [Eq. (\ref{solution})] is shown for different plasma frequencies, panel (a), different electron-ion collision frequencies,  panel (b), different pulse durations,  panel (c), and for different pulse central wavelengths,  panel (d).}}
\label{parametric}
\end{center}
\end{figure}

 {
This analytical model allows us to explain the THz radiation emission in PIC simulations. (i) We show that the current source of the THz emission originates from the electrons which are trapped between the double layers. (ii) The current source is parallel to the incident laser polarization, and consequently, the THz radiation is polarized like the incident laser polarization. (iii) The THz radiation shows a much higher signal along the angles corresponding to the forward direction ($0^{\circ}<\theta\leq90^{\circ}$) than for the backward direction ($90^{\circ}<\theta\leq180^{\circ}$). This is due to the coherence of the phases of the dipole moments induced along the plasma rod [Eq. (\ref{ana_spe_ana})].
 
It would be of interest to see what parameters affect the THz radiation.  The THz radiation forms due to the oscillating dipole moments in the plasma. Hence, the peak frequency of the THz spectrum is determined by plasma frequency ($\varpi^{2}=\Omega_{\rm p}^2-\Gamma^{2}$) as shown in Fig. \ref{parametric}(a).  A higher plasma frequency will cause a greater radiation force and increases the energy of THz radiation. The electron-ion collision frequency $\Gamma$ is one of the important factors affecting the THz spectrum. Increasing the collision time slows down the thermal equilibrium between the electrons and ions and leads to a longer-lasting ambipolar electric field and a broader THz spectrum as shown in Fig. \ref{parametric}(b).  

The THz wave amplitude is proportional to the radiation force driven ${\mathscr{E}_{0}}$ field as expressed in Eq. (\ref{solution}).  To have an estimate of this amplitude, let us suppose a pulse intensity given by $I=I_0/\sqrt{\pi}\exp\left( -{r^2}/{w_0^2}\right)\exp\left(-{t^2}/{T^2}\right)$, where $T$ is the duration of the pulse, and $w_0$ the waist of the pulse. Under the equilibrium between radiation and space charge forces, the THz wave amplitude reads ${\mathscr{E}_{0}}={4E_{\rm p}^2}/{(ec\omega_0^2T^3)}$, where the pulse energy is $E_{\rm p}=I_0{\pi w_0^2 T}$.  Hence, reducing the pulse duration while keeping the pulse energy constant strongly increases the radiated THz energy  [Fig. \ref{parametric}(c)]. This is due to the higher peak intensity of the incident field.  Moreover, increasing the laser wavelength enforces the exerted radiation force on electrons during a laser cycle [Eq. (\ref{radiation_force_3})]. It leads to a stronger net electron current and amplification of the THz radiation [Fig. \ref{parametric}(d)].

}

\section{Discussion}\label{Discussion}

In this work, we have extended our study of femtosecond Bessel beam-induced plasmas inside the dielectrics. A single-shot Bessel beam can generate a high aspect ratio over-critical plasma inside the dielectric.\cite{ardaneh_2021} The generated plasma offers a promising medium for the THz radiation due to the current hot electrons driven by the resonance absorption.

 {
 Based on an analytical approach, we derived the current source, the electric field components, and the angular distribution of THz radiation. The analytical derivation reproduces the main characteristics of the THz radiation calculated using the radiation diagnostic of PIC simulation. Under the linear mode conversion, the radiation force of the resonantly driven plasma waves kicks the electrons from the critical surfaces. Due to the different mobility of the plasma species, charge separations known as double layers, and consequently, ambipolar electric fields form at the plasma surfaces. Most of the ejected electrons from the critical surfaces trap in the potentials of the ambipolar electric fields at plasma edges. The trapped electrons oscillate with a period of around 130~fs and radiate in the THz frequency domain. }

Although in this work we have examined the over-critical plasma, the presented study is valid for the sub-critical plasma, because the radiation force of an intense laser ($\gtrsim10^{18}$~W/cm$^{2}$) can also induce the quasi-static fields and the associated THz radiation. The second-harmonic part of the ambipolar field offers an experimental diagnostic for the detection of THz radiation. Its pattern at the far-field (a central spot) differs from the one generated at the critical surfaces (two lobes parallel with the incident laser polarization discussed in Paper II\cite{Ardaneh_2022b}). 

{ 
To estimate power radiated within the THz range, equating the radiation force with the force due to the space charge field generated from electron-ion separation gives an acceleration $a={e{\mathscr{E}_{0}}}/{m}={4E_{\rm p}^2}/{(mc\omega_0^2T^3)}$, and the power using the Larmor formula\cite{jackson_1998} $P=2e^{2} a^{2}/{3c^{3}}$, is 
$P={32 e^2 E_{\rm p}^2}/{(3m^2 c^5 \omega_0^4T^6)}$, or $P_{\rm W}\approx10^{8}\left[E_{\rm p}^{\rm{\mu J}} (\lambda_0^{\mu\rm{m}})^{2}/(T^{\rm{fs}})^{3}\right]^{2}$.

Assuming a microjoule laser pulse with a duration of 100~fs at 800~nm  wavelength, the laser to THz efficiency is predicted to be about $\sim 10^{-8}$.  This value appears to be very small compared with the THz efficiency of $\sim 10^{-3}-10^{-6}$ for femtosecond pulses with the energy of $\sim 10^{-3}-1$~Joules.\cite{li_2006,li_2014,dechard_2020}  Unlike our works, the absorption process for interactions of  $10^{-3}-1$~Joules class lasers with solids relies on the {Brunel mechanism\cite{chopineau_2019,Ong_2021} }and THz radiation generation is due to the surface currents\cite{li_2006,li_2012} or highly relativistic particles passing through the different dielectrics, the so-called transition radiation.\cite{dechard_2020} The THz energy might be improved using the Bessel beams. The long plasma (recently, we reached cm-scale over-critical plasmas inside dielectrics)\cite{Meyer2019} created by Bessel beams yields a longer double layer at the plasma surface.  A longer double layer traps the ejected hot electrons from the critical surface on a longer distance, for a longer time which results in a longer THz pulse. 

Assuming that individual electrons radiate incoherently, we might estimate the THz intensity and conversion efficiency in the presented PIC simulation. As reported in Paper I\cite{Ardaneh_2022a}, the high-energy electrons represent around 4\% of the electrons in the simulation ($\sim 10^{9}$). Considering the THz intensity for a single electron of $~0.04$~W/cm$^{2}$ [Fig. \ref{pow_spe_fig}(f)], the radiated THz intensity amounts to about $~10^{6}$~W/cm$^{2}$, corresponding to a conversion efficiency of $10^{-8}$, in agreement with the above predicted efficiency.}
 
We require a picosecond timescale to observe the complete process of the THz wave generation (for example, a THz wave at 0.5 THz corresponds to a timescale of 2 ps). However, the numerical heating appearing for high-density plasmas in several picoseconds imposes a limitation on the maximum duration of the PIC simulations.\cite{Arber_2015} For this reason, we did not run our simulations beyond 320~fs. The THz radiation corresponds to a frequency around 30~rad/s. Another challenge is calculating the radiation integral using the whole  set of electrons in the PIC simulation. This requires the implementation of the radiation integral in the MPI-based parallel PIC codes, as done in Refs.\cite{frederiksen_2010,nishikawa_2011} 

\begin{acknowledgments}
We thank the EPOCH support team for help  \url {https://cfsa-pmw.warwick.ac.uk}. The authors acknowledge the financial supports of: European Research Council (ERC) 682032-PULSAR, Region Bourgogne-Franche-Comte and Agence Nationale de la Recherche (EQUIPEX+ SMARTLIGHT platform ANR-21-ESRE-0040), Labex ACTION ANR-11-LABX-0001-01, I-SITE BFC project (contract ANR-15-IDEX-0003), and the EIPHI Graduate School ANR-17-EURE-0002. This work was granted access to the PRACE HPC resources MARCONI-KNL,  MARCONI-M100, and GALILEO at CINECA, Casalecchio di Reno, Italy, under the Project "PULSARPIC" (PRA19\_4980), PRACE HPC resource Joliot-Curie Rome at TGCC, CEA, France under the Project "PULSARPIC" (RA5614), HPC resource Joliot-Curie Rome/SKL/KNL at TGCC, CEA, France under the projects A0070511001 and A0090511001, and M\'{e}socentre de Calcul de Franche-Comt\'{e}.  
\end{acknowledgments}

\appendix
\section{Radiation force density}\label{Radiation force density}
The radiation force per unit volume, force density $\mathbf{f}_{\rm RF}$, on the free electrons can be written:\cite{Landau_1984}
\begin{equation}\label{radiation_force_2}
    {\mathbf{f}_{\rm RF}} =\frac{\epsilon-1}{8\pi}\boldsymbol\nabla  E^{2}+\frac{\epsilon-1}{4\pi c} {\partial_{\rm t}}({\mathbf{E}}\times {\mathbf{B}})
\end{equation}

Usually, the average values of the force density during one period of the laser wave are considered. This is because the time envelope of the laser wave is much slower in comparison with the frequency of the laser wave. Hence, one can neglect the time average of the Poynting term, the last term in Eq. (\ref{radiation_force_2}). Let us consider the monochromatic solutions of the wave equation ${\mathbf{E}}={\mathbf{E}}_{\rm s}(\mathbf{r}) \cos (\omega_0t)$ where ${\mathbf{E}}_{\rm s}(\mathbf{r})$ includes the field's spatial dependence. The radiation force density then reads

\begin{equation}\label{radiation_force_3}
\begin{split}
\mathbf{f}_{\rm{RF}}=&-\frac{\omega_{\rm{pe}}^{2}}{8\pi\omega_0^{2}}\cos ^{2}(\omega_0t)\boldsymbol\nabla{E}_{\rm s}^2\\
=&-\frac{\omega_{\rm{pe}}^{2}}{16\pi\omega_0^{2}}\boldsymbol\nabla{E}_{\rm s}^2
-\frac{\omega_{\rm{pe}}^{2}}{16\pi\omega_0^{2}}\cos(2\omega_0t)\boldsymbol\nabla{E}_{\rm s}^2\\
=&-\frac{\omega_{\rm{pe}}^{2}}{\omega_0^{2}}\boldsymbol\nabla\left\langle\frac{{E}^2}{8\pi}\right\rangle
-\frac{\omega_{\rm{pe}}^{2}}{\omega_0^{2}}\cos(2\omega_0t)\boldsymbol\nabla\left\langle\frac{{E}^2}{8\pi}\right\rangle
\end{split}
\end{equation}

We have used $\left\langle E^{2}\right\rangle={E_{\rm s}^{2}}/{2} $ in Eq. (\ref{radiation_force_3}). 

\section{Ambipolar electric field of double layer}\label{Ambipolar electric field of double layer}
Following Refs.,\cite{lalousis1983,goldsworthy_1986,hora_2008} we use the two-fluid plasma equations for continuity and momentum to derive an analytical solution for the ambipolar electric field of the double layer. The continuity equations read

\begin{subequations}\label{continuty}
\begin{align}
\partial_{\rm t} \left(n_{\rm{e}} m_{\rm{e}}\right)+\partial_{\rm x} \left(n_{\rm{e}}m_{\rm{e}}v_{\rm{e}}\right)=&0 \\
\partial_{\rm t} \left(n_{\rm{i}} m_{\rm{i}}\right)+\partial_{\rm x}\left( n_{\rm{i}}m_{\rm{i}}v_{\rm{i}}\right)=&0
\end{align}
\end{subequations}
where indexes $e$ and $i$ refer to electrons and ions, respectively. The equations for conservation of momentum read:
\begin{subequations}\label{momentum}
\begin{align}
\begin{split}
\partial_{\rm t} \left(n_{\rm{e}}m_{\rm{e}} v_{\rm{e}}\right)=&-\partial_{\rm x} \left(n_{\rm{e}} m_{\rm{e}} v_{\rm{e}}^{2}\right)
-{\partial_{\rm x}P_{\rm{e}}}-en_{\rm{e}}E_{\rm a}\\
&-n_{\rm{e}}m_{\rm{e}}\nu_{\rm ei}\left(v_{\rm{e}}-v_{\rm{i}}\right)+f_{\rm{RF}} 
\end{split}\\
\begin{split}
\partial_{\rm t} \left(n_{\rm{i}}m_{\rm{i}} v_{\rm{i}}\right)=&-\partial_{\rm x} \left(n_{\rm{i}} m_{\rm{i}} v_{\rm{i}}^{2}\right)
-{\partial_{\rm x}P_{\rm{i}}}+en_{\rm{i}}ZE_{\rm a}\\
&+n_{\rm{e}}m_{\rm{e}}\nu_{\rm ei}\left(v_{\rm{e}}-v_{\rm{i}}\right)
\end{split}
\end{align}
\end{subequations}
In Eq. (\ref{momentum}a), the radiation force density is given by Eq. (\ref{radiation_force_3}). We have neglected the radiation force on the ions ${Zm_{\rm e}}/{m_{\rm i}}f_{\rm{RF}}$ in Eq. (\ref{momentum}b).   

The Gauss law for the electric field $E_{\rm a}$ reads:
\begin{equation}
{\partial _{\rm x}}E_{\rm a}=-4\pi e\left(n_{\rm e}-Zn_{\rm i}\right)    
\end{equation}
Taking the time derivative of the Gauss law, using the equations of continuity in Eqs. (\ref{continuty}), and spatial integration gives:
\begin{equation}
    \partial_{\rm t}E_{\rm a}=4 \pi e\left(n_{\rm{e}}v_{\rm{e}}-Zn_{\rm{i}}v_{\rm i}\right)
\end{equation}
The second derivative in time results in:
\begin{equation}
    {\partial_{\rm t}^{2}}E_{\rm a}=4\pi e\left[{\partial_{\rm t}}\left(n_{\rm e} v_{\rm e}\right)-Z{\partial_{\rm t}}\left(n_{\rm i} v_{\rm i}\right)\right]
\end{equation}
Substituting from the equations of momentum in Eqs. (\ref{momentum}) results in

\begin{equation}
    \begin{split}
\frac{1}{4\pi e}{\partial_{\rm t}^{2}}E_{\rm a}=-&{\partial_{\rm x}\left(n_{\rm e} v_{\rm e}^{2}\right)}-\frac{1}{m_{\rm e}} {\partial_{\rm x} P_{x}}-\frac{e n_{\rm e} E_{\rm a}}{m_{\rm e}}\\
+&\nu_{\rm ei} n_{\rm e}\left(v_{\rm i}-v_{\rm e}\right)+\frac{f_{\rm{RF}}}{m_{\rm e}}\\
+Z&{\partial_{\rm x}\left(n_{\rm i}v_{\rm i}^{2}\right)}+\frac{Z}{m_{\rm i}}{\partial_{\rm x} P_{\rm i}}-\frac{Z^{2}en_{\rm i} E_{\rm a}}{m_{\rm i}}\\
+Z&\nu_{\rm ei} n_{\rm e}\left(v_{i}-v_{e}\right)\frac{m_{\rm e}}{m_{\rm i}}
    \end{split}
\end{equation}
The rearrangements of the terms result in the following differential equation that described a damped oscillator subjected to an external force (inhomogeneous second-order differential equation).

\begin{equation}\label{2nd-inhomo}
    {\partial_{\rm t}^{2} E_{\rm a}}+2\Gamma {\partial_{\rm t} E_{\rm a}}+\Omega_{\rm p}^2E_{\rm a}=\Omega_{\rm p}^2\left[\mathscr{E}_{0}+\mathscr{E}_{2}\cos(2\omega_0t)\right]
\end{equation}
where 
\begin{subequations}\label{cons}
\begin{align}
\Gamma=&\frac{\nu_{\rm ei}}{2}\left(1+\frac{Z m_{\rm e}}{m_{\rm i}}\right)\\
\Omega_{\rm p}^2=&\omega_{\rm{pe}}^2\left(1+\frac{Z m_{\rm e}}{m_{\rm
i}}\right)\\
\begin{split}
\mathscr{E}_{0}=&\frac{4 \pi e}{\Omega_{\rm p}^2}\left[{\partial_{\rm x}}\left(Z\frac{P_{\rm i}}{m_{\rm i}}-\frac{P_{\rm e}}{m_{\rm e}} + Zn_{\rm i} v_{\rm i}^{2}-n_{\rm e} v_{\rm e}^{2}\right)\right]\\
&-\frac{4 \pi e}{m_{\rm e}{\Omega_{\rm p}^2}}  \frac{\omega_{\rm{pe}}^{2}}{\omega_0^{2}}\partial_{\rm x}\left\langle\frac{{E}^2}{8\pi}\right\rangle
\end{split}\\
\mathscr{E}_{2} = & -\frac{4 \pi e}{m_{\rm e}{\Omega_{\rm p}^2}}  \frac{\omega_{\rm{pe}}^{2}}{\omega_0^{2}}\partial_{\rm x}\left\langle\frac{{E}^2}{8\pi}\right\rangle
\end{align}
\end{subequations}

\section*{References}
\bibliography{manuscript}
\end{document}